\newif{\ifjournal}
  \journalfalse

\ifjournal
  \documentclass{aa}
  \usepackage{graphicx,times,amssymb}
\else
  \documentclass{paper}
\fi

\renewcommand{\d}{\mathrm{d}}

\begin{document}

\title{Halo Concentrations and Weak-Lensing Number Counts in
  Dark Energy Cosmologies}
\ifjournal
  \author{Matthias Bartelmann\inst{1}, Francesca Perrotta\inst{2,3}
    \and Carlo Baccigalupi\inst{2,4}}
  \institute
   {$^1$ Max-Planck-Institut f\"ur Astrophysik, P.O.~Box 1317,
    D--85741 Garching, Germany\\
    $^2$ Lawrence Berkeley National Laboratory, 1 Cyclotron Road,
    Berkeley, CA 94720, USA\\
    $^3$ Osservatorio Astronomico di Padova, Vicolo dell'Osservatorio
    5, 35122 Padova, Italy\\
    $^4$ SISSA/ISAS, Via Beirut 4, 34014 Trieste, Italy}
  \authorrunning{M.~Bartelmann, F.~Perrotta, C.~Baccigalupi}
  \titlerunning{Halo Concentrations and Weak Lensing in Dark Energy
    Cosmologies}
\else
  \author{Matthias Bartelmann$^1$, Francesca Perrotta$^{2,3}$ and
    Carlo Baccigalupi$^{2,4}$\\
    $^1$ Max-Planck-Institut f\"ur Astrophysik, P.O.~Box 1317,
    D--85741 Garching, Germany\\
    $^2$ Lawrence Berkeley National Laboratory, 1 Cyclotron Road,
    Berkeley, CA 94720, USA\\
    $^3$ Osservatorio Astronomico di Padova, Vicolo dell'Osservatorio
    5, 35122 Padova, Italy\\
    $^4$ SISSA/ISAS, Via Beirut 4, 34014 Trieste, Italy}
\fi

\date{\emph{Astronomy \& Astrophysics, submitted}}

\newcommand{\abstext}
  {We study the effects of a dark energy component with equation of
   state $p=w\rho$ with constant $w\ge -1$ on the formation of Cold
   Dark Matter (CDM) haloes. We find two main effects: first, haloes
   form earlier as $w$ increases, and second, the amplitude of the
   dark-matter power spectrum gets reduced in order to remain
   compatible with the large scale Cosmic Microwave Background (CMB)
   anisotropies. These effects counteract. Using recipes derived from
   numerical simulations, we show that haloes are expected to be up to
   $\sim50\%$ more concentrated in CDM models with quintessence
   compared to $\Lambda$CDM models, the maximum increase being reached
   for $w\sim-0.6$. For larger $w$, the amplitude of the power
   spectrum decreases rapidly and makes expected halo concentrations
   drop. Halo detections through weak gravitational lensing are highly
   sensitive to halo concentrations. We show that weak-lensing halo
   counts with the aperture-mass technique can increase by a factor of
   $\sim2$ as $w$ is increased from $-1$ to $-0.6$, offering a new
   method for constraining the nature of dark energy.}

\ifjournal
  \abstract{\abstext%
    }
\else
  \begin{abstract}
    \abstext
  \end{abstract}
\fi

\maketitle

\section{Introduction}

In recent years, cosmology has seen increasing observational evidence
for an accelerating phase of the cosmic expansion, most notably
through the observations of distant type Ia supernovae (Perlmutter et
al.~1999, Riess et al.~1998). This astonishing evidence motivated
renewed interest in the properties of the energy density ascribed to
the ``vacuum''. A vacuum energy component should account for both the
accelerating expansion and for the residual $\sim70\%$ of the energy
density required for reconciling the geometrical flatness required by
Cosmic Microwave Background (CMB) observations (De Bernardis et
al.~2002, Lee et al.~2001, Halverson et al.~2002) with the evidence of
a low-density universe with $\Omega_0\sim0.3$ (Percival et al.~2001).

While one of the historical candidates for such an energy density is
the cosmological constant, introduced as a simple geometrical term in
Einstein's equations, it is well known that it leads to serious and
unsolved theoretical problems. The exceedingly low value of the vacuum
energy density today, compared to that allowed by the most plausible
theories of the early stage of the Universe, motivated the
introduction of a more general concept now widely known as ``dark
energy''.

Preceding the evidence for cosmic acceleration, a generalisation of
the cosmological constant by means of a scalar field, now known as
``quintessence'', was proposed (Wetterich 1988, Ratra \& Peebles
1988). In this class of models, the dark energy is supposed to reside
mostly in the potential energy of the field, which interacts only
gravitationally with ordinary matter. The evolution is described by
the ordinary Klein-Gordon equation. If the potential is flat enough,
or if the motion of the field along its trajectory is sufficiently
slow, a cosmological constant-like behaviour can be mimicked by the
scaling of the energy density of the quintessence field.

For general forms of the scalar field potential, there exist attractor
trajectories for the evolution of the background expectation value of
the field. These trajectories are known as ``tracking'' (Steinhardt,
Wang \& Zlatev 1999) and ``scaling'' (Liddle \& Scherrer 1999)
solutions. They have been shown to alleviate, at least at a classical
level, the fine-tuning required in the early Universe, when the
typical energy scales were presumably comparable to the Planck scale,
to generate a vanishing relic vacuum energy as it is observed today,
120 orders of magnitude smaller. However, these scenarios are not able
to solve the coincidence problem, i.e. why we are living in the epoch
in which dark energy and matter have roughly the same energy
density. Despite some attempts at addressing this issue
(Tocchini-Valentini \& Amendola 2002, Chiba 2001, Armendariz-Picon,
Mukhanov \& Steinhardt 2001, Dodelson, Kaplinghat \& Stewart 2000), it
remains one of the greatest puzzles of modern cosmology.

For constraining the nature of the dark energy, an important step
would be accomplished if parameters could be constrained which capture
its most essential features. In particular, if the dark energy is
modelled as a quintessence field in the tracking regime, the simplest
description of its properties will require the use of only two
parameters, i.e.~its present energy density and the ratio between
pressure and energy density in its equation of state. Generally, this
ratio depends on time, as implied by the evolution according to the
equation of motion. However, it can be shown that, in most simple
models of quintessence involving an inverse power-law potential, the
effect of a time variation of the equation of state can be neglected
at low redshifts, when the field has settled on its tracking
trajectory. In this case, the equation of state is simply related to
the power with which the potential depends on the field itself. This
simplification allows constructing a scheme for describing the dark
energy behaviour at redshifts where interesting cosmological effects
arise, such as the effect on the magnitude-redshift relation of type
Ia supernovae (Perlmutter 1999, Riess 1998) and the effect on
gravitational lensing of distant galaxies and quasars (Futamase \&
Yoshida 2001).

Even though the dark energy dynamics has a geometrical effect on
acoustic features of the CMB anisotropy (Baccigalupi et al.~2002;
Doran, Lilley \& Wetterich 2002; Corasaniti \& Copeland 2002), it is
now commonly accepted that the most interesting properties of a dark
energy component have to be probed by looking at processes occurring
at low redshifts when it starts dominating the cosmic expansion.

Thus, one of the most powerful probes of the quintessence field
results from its effects on cosmic structure formation, most notably
on the background cosmology and the evolution of individual collapsing
overdensities. First, a dark energy component affects the matter
density of the background in which haloes form. Second, the amplitude
of matter perturbations is sensitive to the presence of a dynamical
vacuum energy, through the normalisation of the matter power spectrum
to the large, unprocessed, scales probed by the large-scale CMB
anisotropies. Third, changes in the background matter density induced
by a dark energy component can seriously affect characteristic
properties of collapsing structures. As shown by {\L}okas \& Hoffmann
(2002), a substantial quintessence component changes the
characteristic density of a forming dark matter halo.

In this paper, we study how quintessence affects the concentration of
dark-matter haloes, and resulting changes in their weak-lensing
efficiency. Weak gravitational lensing provides a powerful tool for
mapping the large-scale mass distribution (see Mellier 1999a and
Bartelmann \& Schneider 2001 for reviews), and the potential impact of
dark energy on the weak lensing convergence power spectrum has already
been recognised (see Mellier 1999b; Huterer 2002). We show in this
paper that an energy density component with negative pressure affects
weak lensing by dark-matter haloes not only through changes in the
global properties of the Universe, but also by modifying their
internal density concentration. The main idea is that dark energy
affects structure growth and thus the time of halo formation. Since
halo concentrations reflect the density of the Universe at their
formation epoch, this affects halo mass distributions, and weak
lensing provides methods for quantifying respective changes.

The paper is organised as follows. In Sect.~2, we describe the main
effects of the dark energy on the cosmological growth factor, volume
elements, and the normalisation of the dark-matter power spectrum. In
Sect.~3 we compute the impact on halo concentration. In Sect.~4 we
predict resulting effects on the weak-lensing aperture mass
statistics, and Sect.~5 contains our conclusions.

\section{Cosmological implications of quintessence models}

We model the dark energy as a spatially homogeneous component,
labelled $Q$, with constant equation of state parameterised by the
ratio between the pressure $p_\mathrm{Q}$ and the energy density
$\rho_\mathrm{Q}$, $w=p_\mathrm{Q}/\rho_\mathrm{Q}$. We neglect a
possible time variation of $w$, as well as any effects possibly due to
spatial inhomogeneities of the dark energy. Indeed, at least in most
models proposed so far, the relevant cosmological effects of the dark
energy compared to a cosmological constant are mainly related to its
effective equation of state at redshifts when it is relevant for
cosmic expansion, say $z\le5$, as we already noted in the
introduction. Neglecting inhomogeneities of the dark energy is
justified in the present context since they are likely to show
relativistic behaviour on sub-horizon cosmological scales; indeed, the
effective mass of the vacuum component, of the order of the critical
density today, is extremely light compared to any other known massive
particle, so that quintessence clustering occurs only on scales larger
than or equal to the horizon size (Ma et al.~1999). It can also be
shown formally that a minimally coupled quintessence field has a
relativistic effective sound speed (Hu 1998), so that its fluctuations
are damped out on the scales in which we are interested here.

Assuming the equation of state $p_\mathrm{Q}=w\rho_\mathrm{Q}$, the
adiabatic equation implies
\begin{equation}
  \rho_\mathrm{Q}=\rho_\mathrm{Q,0}\,a^{-3(1+w)}\;,
\label{eq:1}
\end{equation}
where $\rho_\mathrm{Q,0}$ is the quintessence energy density today,
and $a$ is the cosmic scale factor normalised to unity at the present
epoch. Friedmann's equation can then be written,
\begin{equation}
  H^2(t)=H_0^2\,\left[\Omega_\mathrm{Q}a^{-3(1+w)}+\Omega_0a^{-3}+
  (1-\Omega_\mathrm{Q}-\Omega_0)a^{-2}\right]\;,
\label{eq:2}
\end{equation}
where $\Omega_\mathrm{Q}$ is the quintessence density parameter at the
present epoch, i.e.~$\rho_\mathrm{Q,0}$ divided by the critical energy
density today. The density parameter for non-relativistic matter is
$\Omega_0$ today. Obviously, for $w=-1$, Friedmann's equation for a
cosmological constant is retained. Unless stated otherwise, we will
assume in this paper that the curvature of spatial hypersurfaces is
zero, thus $\Omega_\mathrm{Q}+\Omega_0=1$, and the curvature term in
(\ref{eq:2}) vanishes.

The quintessence term in (\ref{eq:2}) has two immediate consequences
relevant for our purposes. First, the way how density inhomogeneities
grow is modified, and second, the cosmic volume per unit redshift
changes.

\begin{figure}[ht]
  \includegraphics[width=\hsize]{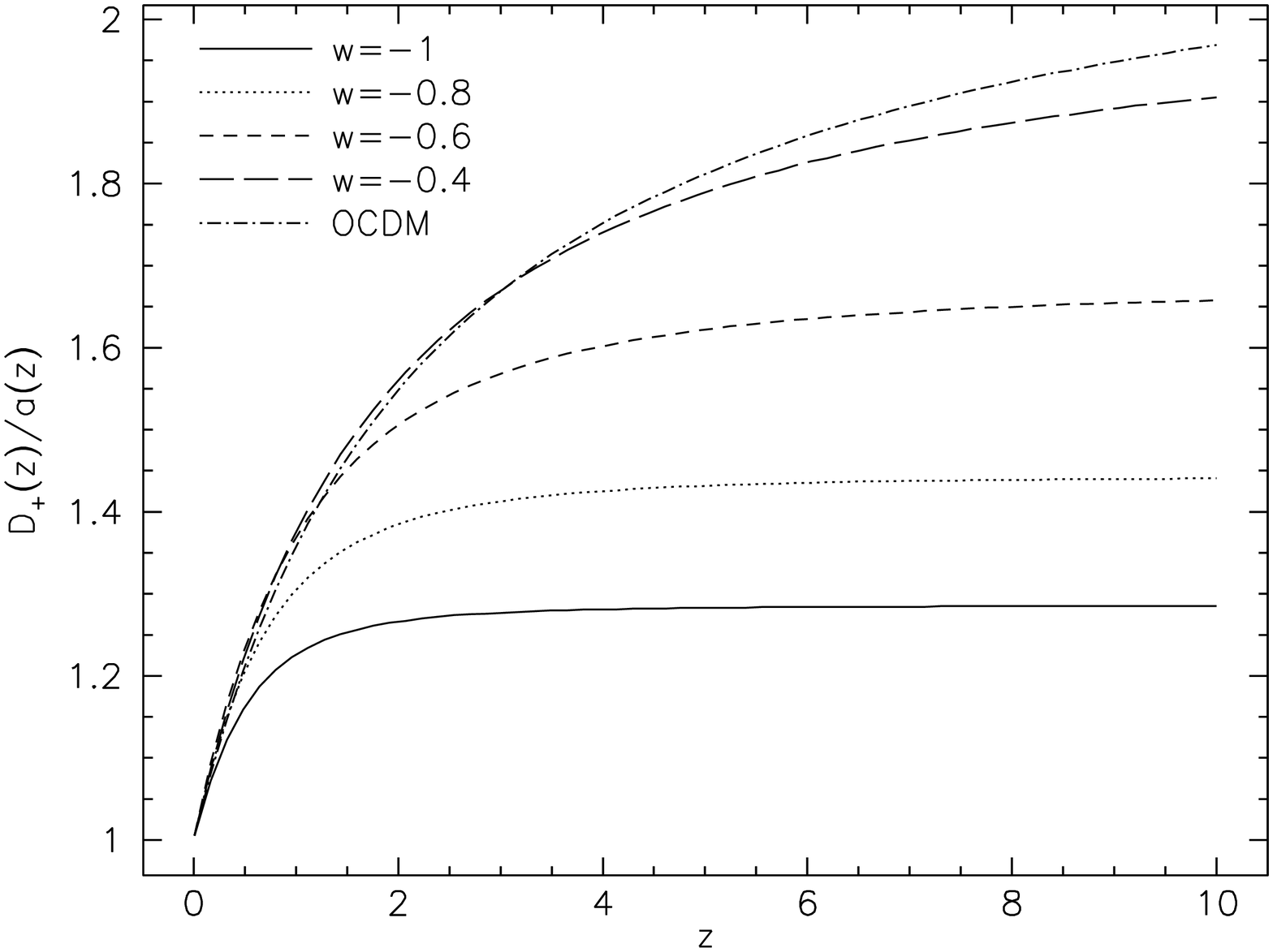}
\caption{Growth factor $D_+(z)$ as a function of redshift for five
  different cosmological models as indicated. The growth factor is
  normalised to unity at the present epoch, and divided by the scale
  factor to emphasise the differences between the models. With
  increasing $w$, the growth factor increases towards its value for
  the open model with $\Omega_\mathrm{Q}=0$.}
\label{fig:1}
\end{figure}

In linear theory, the density contrast $\delta$ of matter
perturbations grows according to
\begin{equation}
  \delta\propto\frac{\dot{a}}{a}\int_0^a\,\frac{\d a}{\dot{a}^3}
\label{eq:3}
\end{equation}
(Heath 1977; see also chapter 15 of Peacock 1999). The right-hand side
of (\ref{eq:3}) is proportional to the so-called growth factor
$D_+(a)$, which is commonly normalised either to unity at $a=1$, or
such that it rises proportional to $a$ for $a\ll1$. We plot in
Fig.~\ref{fig:1} the growth factor as a function of redshift $z$,
normalised to unity today and divided by $a$. The normalisation
ensures that $D_+/a$ goes to unity for $z\to0$. In an Einstein-de
Sitter model, $D_+=a$, thus the curves show how much faster structures
grow in the assumed model universes compared to an Einstein-de Sitter
model. Curves are shown for a variety of cosmological models. All of
them have $\Omega_0=0.3$ and either $\Omega_\mathrm{Q}=1-\Omega_0$ or,
for comparison, $\Omega_\mathrm{Q}=0$.

The solid curve in Fig.~\ref{fig:1} shows the growth factor for a
model universe with a cosmological constant ($w=-1$). Going back in
time, $D_+/a$ rises from unity to $\sim1.3$ and turns flat near
$z\sim\Omega_0^{-1}\sim3$. This means that structures start forming
more quickly in this model compared to an Einstein-de Sitter model,
but the growth slows down considerably at redshifts smaller than
$z\sim3$. For the low-density open model without quintessence
(labelled $\Omega_\mathrm{Q}=0$), $D_+/a$ keeps rising as $z$
increases. At redshift $5$, for instance, the amplitude of structures
is $1.4$ times higher than in the cosmological-constant
model. Increasing $w$ interpolates between the open and the flat model
with cosmological constant. Thus, keeping $\Omega_0$ and
$\Omega_\mathrm{Q}$ fixed, structures form earlier for larger values
of $w$, approaching the growth behaviour for low-density open models
without quintessence or cosmological constant.

\begin{figure}[ht]
  \includegraphics[width=\hsize]{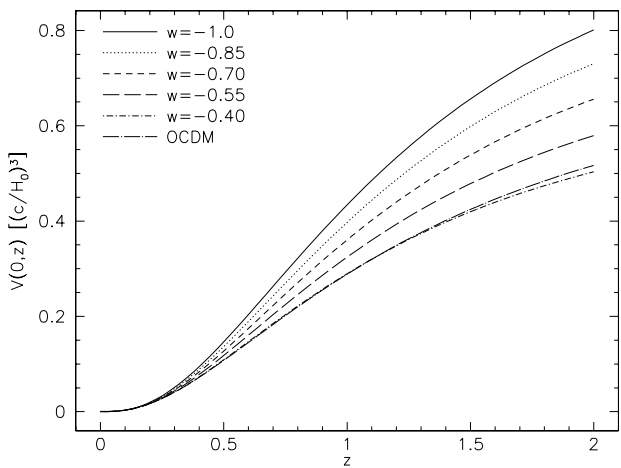}
\caption{Cosmic volume in units of the Hubble volume between redshift
  zero and $z$ for six different cosmological models as
  indicated. With increasing $w$, the cosmic volume decreases towards
  its value for the open model with $\Omega_\mathrm{Q}=0$.}
\label{fig:2}
\end{figure}

A similar interpolation is seen in the behaviour of the cosmic volume
per unit redshift. Figure~\ref{fig:2} shows
\begin{equation}
  V(z)=\int_0^z\d z'\,4\pi D^2(z')\,
  \left|\frac{c\d t}{\d z}\right|(z')\;,
\label{eq:4}
\end{equation}
where $D(z)$ is the angular-diameter distance between redshifts $0$
and $z$, thus $V(z)$ is the proper volume of a sphere of ``radius''
$z$ around the observer. Similar to the behaviour of the growth
factor, dark-energy models interpolate between the two limiting
curves, where the model with cosmological constant has the largest and
the model without cosmological constant has the smallest volume.

Here and below, we keep the cosmological parameters fixed at
$\Omega_0=0.3$, $\Omega_\mathrm{Q}=0.7$. The Hubble constant is
$H_0=100\,h\,\mathrm{km\,s^{-1}\,Mpc^{-1}}$ with $h=0.7$.

In order to describe the formation of dark-matter haloes, we need to
specify the power spectrum of dark-matter fluctuations. We choose the
cold dark matter (CDM) model, whose transfer function was given by
Bardeen et al.~(1986), and we set the index of the primordial power
spectrum to the Harrison-Zel'dovich value of $n=1$. Then, the power
spectrum has two free parameters, the shape parameter $\Gamma$ which
locates its maximum, and the amplitude. For the shape parameter, we
assume $\Gamma=\Omega_0h$ as suggested by theory and in agreement with
observations of the galaxy power spectrum. For our model for dark
energy, the transfer function by Bardeen et al.~is applicable because
dark energy does not cluster on the relevant scales, and the shape
parameter $\Gamma$ is set by the scale of matter-radiation equality,
which is unaffected by $w$ (cf.~Wang \& Steinhardt 1998).

The amplitude of the power spectrum needs to be chosen such that
certain observations can be reproduced. It is an important constraint
that the abundance of rich galaxy clusters in our cosmic neighbourhood
be reproduced. Since the galaxy-cluster mass function falls very
steeply at the high-mass end, small changes in the amplitude of the
power spectrum lead to large changes in the cluster number density,
thus the amplitude is in principle well constrained by the cluster
abundance. However, for that normalisation procedure, rich galaxy
clusters are identified by their X-ray emission, thus the reliability
of the normalisation depends on how well models can describe the X-ray
properties of the intracluster gas.

\begin{figure}[ht]
  \includegraphics[width=\hsize]{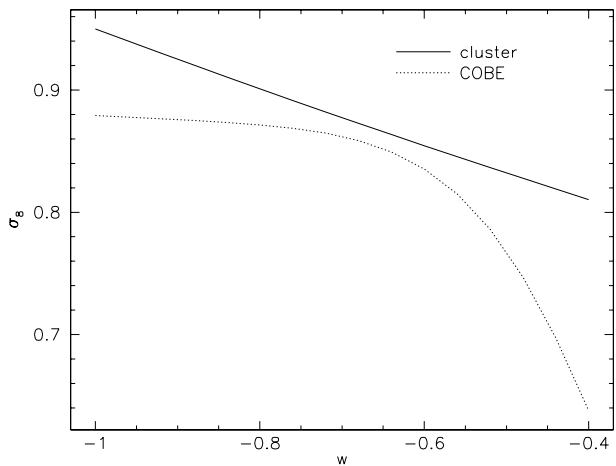}
\caption{The normalisation of the power spectrum, expressed in terms
  of $\sigma_8$, is shown as a function of $w$ for two different
  normalisation methods. The solid line shows the cluster-abundance
  normalisation derived by Wang \& Steinhardt, the dotted line shows
  $\sigma_8$ as constrained by the COBE-DMR data. While $\sigma_8$ is
  almost constant in the latter case for $w\lesssim-0.6$, it drops
  rapidly for larger $w$ because of the increasingly strong integrated
  Sachs-Wolfe effect. The uncertainty of $\sigma_8$ determined from
  COBE data is approximately 7\% (Bunn \& White 1997).}
\label{fig:3}
\end{figure}

We saw before that structure starts forming earlier in quintessence
models with $w>-1$ compared to models with cosmological
constant. Galaxy clusters forming earlier are hotter because of the
higher mean density of their surroundings. Reproducing the current
number density of X-ray selected clusters thus requires a
power-spectrum amplitude which decreases with increasing $w$ (Wang \&
Steinhardt 1998). Figure~\ref{fig:3} shows their results, expressing
the power-spectrum amplitude in terms of the \emph{rms} fluctuation
amplitude $\sigma_8$ on a physical scale of $8\,h^{-1}\mathrm{Mpc}$.

The power spectrum of density perturbations can be normalised by the
CMB anisotropies on large angular scales, corresponding to physical
scales which were larger than the horizon at decoupling. The only
available data on those scales are those of the Differential Microwave
Radiometer (DMR) experiment on board the COsmic Background Explorer
satellite (COBE, Bennett et al.~1996), but much improved data are
being taken by the Microwave Anisotropy Probe (MAP,
http://map.gsfc.nasa.gov/), and will be taken by the Planck satellite
(http://astro.estec.esa.nl/SA-general/Projects/Planck/). Sub-degree
CMB anisotropies have been measured by several experiments (De
Bernardis et al.~2001, Lee et al.~2001, Halverson et
al.~2002). However, on these angular scales, the main effect of
increasing $w$ above $-1$ is to rigidly shift acoustic peaks toward
larger angular scales (see Baccigalupi et al.~2002 and references
therein), having almost no effect on the overall normalisation of the
spectrum.

Normalising the power spectrum according to the COBE measurements
fixes the amplitude of the power spectrum on its large-scale end. As
$w$ increases at fixed $\Omega_0$ and $\Omega_\mathrm{Q}$, dark energy
dominates cosmic expansion earlier compared to a model with
cosmological constant having the same energy density today. This has
the effect of enhancing the dynamics of the gravitational potential
due to the change in the cosmic equation of state, thus increasing the
Integrated Sachs-Wolfe (ISW) effect on COBE scales. While this trend
is gentle for $w\lesssim-0.6$, it steepens for larger $w$, as shown by
the dotted line in Fig.~\ref{fig:3}.

This non-linear behaviour of $\sigma_8$ reflects the sensitivity of
the ISW effect (and thus the normalisation) to the redshift
$z_\mathrm{Q}$ at which dark energy starts dominating over matter. It
is easy to show that this is determined by the power law
$1+z_\mathrm{Q}=(\Omega_0/\Omega_\mathrm{Q})^{-1/3w}$. We recall that,
although we consider values of $w$ up to $-0.4$ for the sake of
generality, values larger than $-0.6$ are not interesting in the
framework of dark-energy models. Indeed, $w$ larger than $-0.6$ is
insufficient to provide cosmic acceleration if $\Omega_\mathrm{Q}=0.7$
as we assume in this work. In addition, the sharp decrease in
Fig.~\ref{fig:3} due to the ISW leads to a decrease of the acoustic
peaks in the CMB power spectrum below the level observed by
experiments operating on sub-degree angular scales.

In order to avoid the uncertainties in $\sigma_8$ due to the
uncertainties in modelling the X-ray cluster population, we choose the
COBE normalisation for our study. We adopt the normalisation method by
Bunn \& White (1997), which has a $7\%$ accuracy. This procedure
exploits a maximum likelihood approach for reproducing the measured
CMB anisotropy power once the sky regions affected by the Galactic
signal have been cut out. The CMB anisotropy can be expressed as a
line-of-sight integral of the perturbation power spectrum weighted
with suitable geometrical functions, implemented in the CMBFAST code
(Seljak \& Zaldarriaga 1996). For taking the dark energy component
into account, we use a modified version of CMBFAST (see Baccigalupi,
Matarrese \& Perrotta 2000). Since halo properties are determined by
the small-scale end of power spectrum while the COBE normalisation
fixes its large-scale end, quantitative results will sensitively
depend on the index $n$ of the power spectrum.

An entirely different and perhaps more direct way of normalising the
power spectrum has become feasible in the recent past. Large-scale
density fluctuations differentially deflect light on its way from
distant sources to us. The gravitational tidal field of the matter
inhomogeneities coherently distorts the images of faint background
galaxies. Albeit weak, this \emph{cosmic shear} effect has recently
been measured successfully by several groups, whose results agree
impressively although different telescopes, observational parameters
and data-reduction techniques were used (Van Waerbeke et al.~2000;
Bacon, Refregier \& Ellis 2000; Kaiser, Wilson \& Luppino 2000;
Wittman et al.~2000; Maoli et al.~2001; Van Waerbeke et
al.~2001). Since gravitational lensing depends only on the matter
distribution and not on its composition or physical state, cosmic
shear should provide one of the cleanest ways for constraining the
power-spectrum amplitude.

\begin{figure}[ht]
  \includegraphics[width=\hsize]{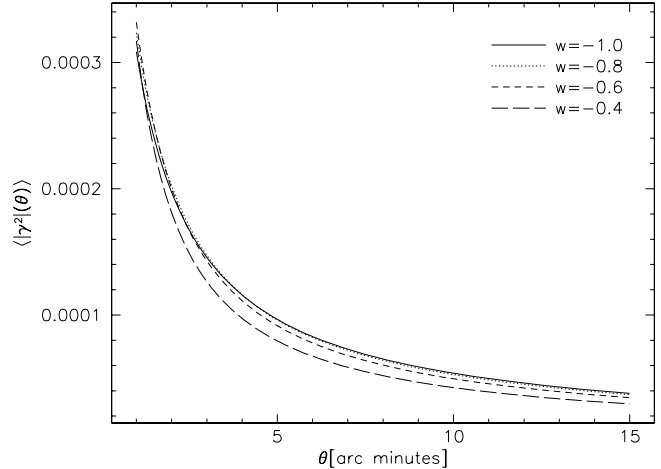}
\caption{The squared cosmic shear $\langle|\gamma^2|\rangle(\theta)$,
  averaged in apertures of radius $\theta$, is shown as a function of
  $\theta$ for five different cosmological models as indicated. A
  source redshift distribution with mean redshift $\sim0.9$ was
  assumed for the plot. Changing $w$ has very little effect on the
  curves, implying that the constraints on $\sigma_8$ derived from
  cosmic-shear measurements are insensitive to $w$.}
\label{fig:4}
\end{figure}

Cosmic-shear measurements have been shown to agree very well with the
expectations in a $\Lambda$CDM universe. Fixing the shape parameter to
$\Gamma=0.21$, the measurements require $\Omega_0<0.4$ and
$\sigma_8>0.7$ at 95\% confidence (Van Waerbeke et al.~2001). Note,
however, that these constraints depend on the nonlinear evolution of
the power spectrum which has some uncertainties.

This result is almost completely insensitive to $w$ in quintessence
models. In Fig.~\ref{fig:4}, we show the \emph{rms} cosmic shear in
apertures of radius $\theta$ as a function of $\theta$ for four models
differing by $w$, as indicated. The power spectrum was normalised to
reproduce the COBE measurements, and its nonlinear evolution was
approximated using the fitting formulae by Peacock \& Dodds (1996).
Following Van Waerbeke et al.~(2001), we adopted the source redshift
distribution shown in (\ref{eq:7a}) below with parameters $z_0=0.8$
and $\beta=1.5$. The curves are much closer than the typical
uncertainty of cosmic-shear measurements, showing that the constraint
on $\sigma_8$ derived from cosmic shear is independent of $w$. The
main reason is that cosmic shear is most sensitive to structures below
redshift $\sim0.5$ where the differences between the models are
small. The figure shows that the COBE normalisation is not in conflict
with the measured cosmic shear for all interesting values of $w$.

\section{Halo concentrations in dark energy models}

Numerical simulations of cosmic structure formation show consistently
that the density profiles of dark-matter haloes can be described by a
two-parameter family of models. Far outside a scale radius
$r_\mathrm{s}$, the profiles fall off proportional to $r^{-3}$, while
they are cuspy but considerably flatter well within
$r_\mathrm{s}$. The exact inner profile slope is under debate. The
second free parameter besides the scale radius is a characteristic
density scale $\rho_\mathrm{s}$. We adopt the density profile
suggested by Navarro et al.~(1995),
\begin{equation}
  \rho(r)=\frac{\rho_\mathrm{s}}
               {(r/r_\mathrm{s})(1+r/r_\mathrm{s})^2}\;.
\label{eq:5}
\end{equation}

The two parameters $r_\mathrm{s}$ and $\rho_\mathrm{s}$ are not
independent. Haloes are commonly parameterised by their virial mass
$M$. For a given cosmology, this also defines their virial radius
$R$. Numerical simulations show that the scale radius depends on the
mass such that the halo concentration $c=R/r_\mathrm{s}$ is a
characteristic function of mass. Given $M$, $R$ and $c$, the density
scale $\rho_\mathrm{s}$ is fixed. For definiteness, we parameterise
halo masses consistently by $M_{200}$, i.e.~masses enclosed in spheres
with radius $R_{200}$ in which the average density is 200 times the
critical density.

Numerically simulated haloes turn out to be the more concentrated the
earlier they form. This is interpreted assuming that the central
density of a halo reflects the mean cosmic density at the time when
the halo formed. This implies that haloes forming earlier are expected
to be more concentrated.

Several algorithms based on this assumption have been suggested for
describing the concentration of dark-matter haloes. Originally,
Navarro et al.~(1997) devised the following approach. A halo of mass
$M$ is first assigned a collapse redshift $z_\mathrm{coll}$ defined as
the redshift at which half of the final halo mass is contained in
progenitors more massive than a fraction $f_\mathrm{NFW}$ of the final
mass. Then, the density scale of the halo is assumed to be some factor
$C$ times the mean cosmic density at the collapse redshift. They
recommended setting $f_\mathrm{NFW}=0.01$ and $C=3\times10^3$ because
their numerically determined halo concentrations were well fit
assuming these values.

Bullock et al.~(2001) noticed that halo concentrations change more
rapidly with halo redshift than the approach by Navarro et al.~(1997)
predicts. They suggested a somewhat simpler algorithm. Haloes are
assigned a collapse redshift defined such that the non-linear mass
scale at that redshift is a fraction $f_\mathrm{B}$ of the final halo
mass. The halo concentration is then assumed to be a factor $K$ times
the ratio of the scale factors at the redshift when the halo is
identified and at the collapse redshift. Comparing with numerical
simulations, they found $f_\mathrm{B}=0.01$ and $K=4$.

Yet another algorithm was suggested by Eke et al.~(2001). They
assigned the collapse redshift to a halo of mass $M$ by requiring that
the suitably defined amplitude of the linearly evolving power spectrum
at the mass scale $M$ equals a constant
$C_\mathrm{ENS}^{-1}$. Numerical results are well represented setting
$C_\mathrm{ENS}=28$.

\begin{figure}[ht]
  \includegraphics[width=\hsize]{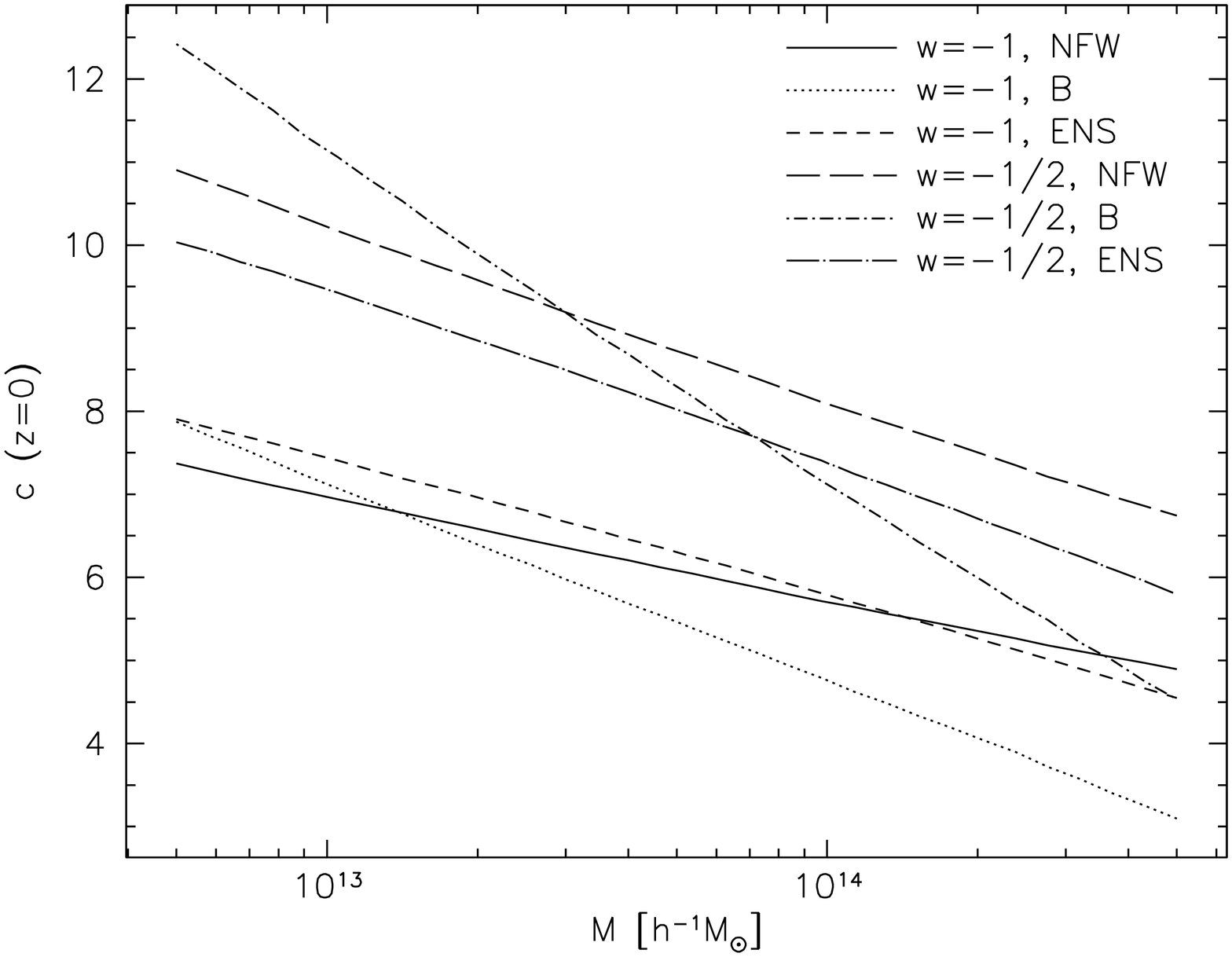}
\caption{Halo concentration parameters $c$ as functions of halo mass
  $M_{200}$. Results obtained from three different prescriptions for
  calculating the concentration are shown; these are the prescriptions
  from Navarro, Frenk \& White (NFW), Bullock et al.~(B) and Eke,
  Navarro \& Steinmetz (ENS). Curves are shown for the $\Lambda$CDM
  model ($w=-1$) and for a quintessence model with
  $w=-1/2$. Irrespective of the prescription, the concentration
  parameters are higher in the quintessence model than in the
  $\Lambda$CDM model for all halo masses.}
\label{fig:5}
\end{figure}

Since we consistently use $R_{200}$ and $M_{200}$ for parameterising
haloes, we need to convert masses and concentration parameters from
the slightly different definitions introduced by Bullock et al.~and
Eke et al. In particular, this requires us to iteratively compute the
concentration parameter according to our definition. Moreover, the
algorithms implicitly use the mean overdensity of virialised haloes,
$\Delta_\mathrm{v}$, and the linear overdensity of collapsed haloes,
$\delta_\mathrm{c}$. These parameters depend on cosmology and on the
dark energy parameter $w$. We compute them using the formulae given in
{\L}okas \& Hoffmann (2002).

Although halo concentrations produced by these different algorithms
differ in detail, they have in common that haloes forming earlier are
more concentrated. Taken together with the earlier result that haloes
form earlier in dark energy models with $w>-1$ than in $\Lambda$CDM
models, this implies that haloes are expected to be more concentrated
in models with $w>-1$. Figure~\ref{fig:5} illustrates this. Halo
concentrations computed with the three different algorithms are
plotted as functions of halo mass for redshift zero. Since less
massive haloes start forming earlier than more massive ones in
hierarchical models like CDM, the concentration decreases with halo
mass. Curves are shown for models with $w=-1$ and $w=-1/2$, keeping
all other parameters fixed. At $M\sim5\times10^{13}\,h^{-1}M_\odot$,
for instance, halo concentrations are approximately 50\% higher in the
dark energy compared to the $\Lambda$CDM model.

Although concentrations of haloes identified at $z>0$ tend to be
smaller than at redshift zero, the trend of increasing concentrations
with increasing $w$ remains also at higher redshifts.

\begin{figure}[ht]
  \includegraphics[width=\hsize]{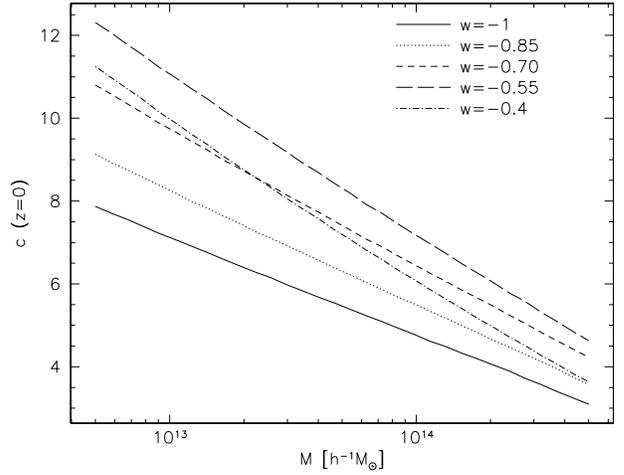}
\caption{Halo concentrations obtained from the algorithm described by
  Bullock et al. as functions of halo mass $M_{200}$ for five
  different cosmological models as indicated. With increasing $w$, the
  concentration increases until $w\gtrsim-0.6$, and drops rapidly as
  $w$ increases further.}
\label{fig:7}
\end{figure}

Figure~\ref{fig:7} shows halo concentrations as functions of mass
derived from the algorithm suggested by Bullock et al. for five
different choices of $w$, keeping all other parameters fixed. As $w$
increases away from $-1$, concentrations first increase for all halo
masses shown. A maximum is reached for $w\sim-0.6$. As $w$ is raised
further, concentrations decrease rapidly. This is an effect of the
power-spectrum normalisation. As $w$ increases, $\sigma_8$ must
decrease because otherwise the normalisation constraints would be
violated; either, there would be too many hot galaxy clusters, or the
secondary CMB anisotropies caused by the integrated Sachs-Wolfe effect
would exceed the COBE measurements. As $\sigma_8$ is lowered, haloes
form later, thus counter-acting the earlier increase of the growth
factor in quintessence models. Adopting the COBE normalisation, the
maximum effect is achieved just before the integrated Sachs-Wolfe
effect sets in strongly at $w\gtrsim-0.6$, cf.~Fig.~\ref{fig:3}.

\section{Weak lensing and halo counts}

The impact of dark energy on halo concentrations should cause an
impact on many observable quantities. As an illustration, we will now
describe how the number of haloes detectable through their weak
gravitational lensing effect change with $w$.

The gravitational tidal field of individual sufficiently massive
haloes imprints coherent distortions on the images of faint background
galaxies in their neighbourhood. Haloes can thus be detected searching
for their characteristic signature on the appearance of the background
galaxy population. A sensitive and convenient technique for halo
detection, the aperture mass technique, has been suggested by
Schneider (1996).

Consider a circular aperture of angular radius $\theta$. The aperture
mass is defined as a weighted integral of the lensing convergence
$\kappa$ across the aperture,
\begin{equation}
  M_\mathrm{ap}(\theta)=\int\d^2\vartheta\kappa(\vec\vartheta)\,
  U(|\vec\vartheta|)\;.
\label{eq:6}
\end{equation}
The convergence is the surface mass density scaled by its critical
value for strong lensing,
\begin{equation}
  \kappa=\frac{\Sigma}{\Sigma_\mathrm{cr}}\;,\quad
  \Sigma_\mathrm{cr}=\frac{c^2}{4\pi G}
  \frac{D_\mathrm{s}}{D_\mathrm{d}D_\mathrm{ds}}\;,
\label{eq:7}
\end{equation}
where $D_\mathrm{d,ds,s}$ are the angular-diameter distance from the
observer to the lens, the lens to the source, and the observer to the
source, respectively.

Through $\Sigma_\mathrm{cr}$, the aperture mass (\ref{eq:6}) depends
on the source redshift. We compute mean aperture masses by averaging
$M_\mathrm{ap}$ over the normalised source-redshift distribution
\begin{equation}
  p(z_\mathrm{s})=\frac{\beta}{z_0^3\,\Gamma(3/\beta)}\,
  \exp\left[-\left(\frac{z}{z_0}\right)^\beta\right]
\label{eq:7a}
\end{equation}
with $z_0=1$ and $\beta=1.5$, implying a mean source redshift of
$\sim1.5$ (cf.~Smail et al.~1995; Cohen et al.~2000).

If the weight function $U(|\vec\vartheta|)$ is compensated,
\begin{equation}
  \int\d^2\vartheta\,U(|\vec\vartheta|)=0\;,
\label{eq:8}
\end{equation}
the aperture mass $M_\mathrm{ap}$ can be written as a (differently)
weighted integral across the aperture of the tangential component of
the shear with respect to the aperture centre. Thus, $M_\mathrm{ap}$
is a directly observable quantity.

Halo detection can then proceed as follows. An aperture of given
radius is shifted across a wide and sufficiently deeply observed
field. At all aperture positions, the aperture mass is
determined. Potential haloes are located where $M_\mathrm{ap}$ exceeds
a certain threshold.

The signal-to-noise ratio of an aperture-mass measurement is given by
$\mathcal{S}=M_\mathrm{ap}\sigma_\mathrm{M}^{-1}$ with the dispersion
\begin{equation}
  \sigma_\mathrm{M}(\theta)=0.016\,
  \left(\frac{n_\mathrm{g}}{30\,\mathrm{arcmin}^2}\right)^{-1/2}
  \left(\frac{\sigma_\epsilon}{0.2}\right)
  \left(\frac{\theta}{1'}\right)^{-1}\;,
\label{eq:9}
\end{equation}
where $n_\mathrm{g}$ is the number density of faint background
galaxies and $\sigma_\epsilon$ is the variance of their intrinsic
ellipticity distribution. We assume $n_\mathrm{g}=30$ and take into
account that only such sources from the distribution (\ref{eq:7a})
contribute to the signal whose redshift is larger than that of the
lensing halo.

Convolving the projected NFW density profile (cf.~Bartelmann 1996)
with the weight function $U(|\vec\vartheta|)$, the aperture mass of
NFW haloes as a function of halo mass, $M_\mathrm{ap}(M,\theta)$ is
easily computed. Assuming further a halo mass function
$N_\mathrm{halo}(M,z)$, we can calculate the number of haloes per unit
mass and redshift whose aperture mass is sufficiently high for the
signal-to-noise ratio $\mathcal{S}$ to exceed a given threshold
$\mathcal{S}_\mathrm{min}$. We choose the mass function suggested by
Sheth \& Tormen (1999), which is a variant of the Press-Schechter
(1974) mass function which well reproduces the mass function found in
numerical simulations. We take into account that our definition of
mass differs slightly from Sheth \& Tormen's in that we use the mass
enclosed by a sphere in which the mean density is 200 times the
\emph{critical} rather than the \emph{mean} density.

\begin{figure}[ht]
  \includegraphics[width=\hsize]{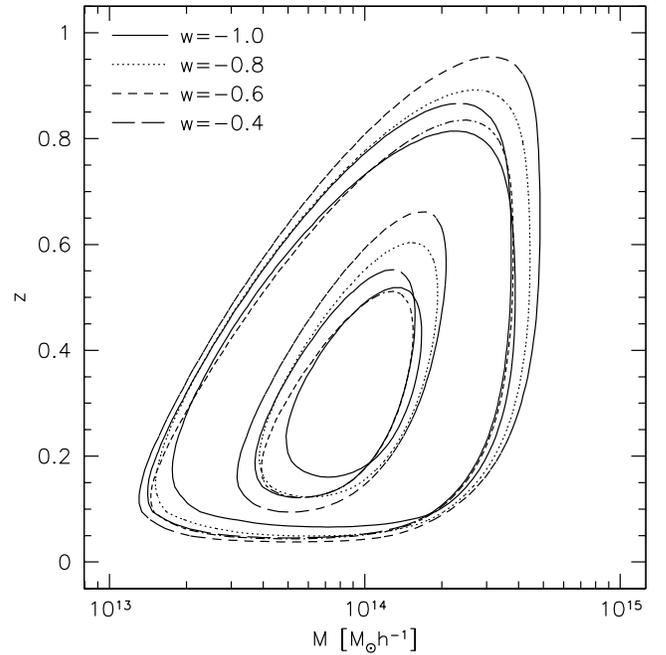}
\caption{Contours in the mass-redshift plane showing the number
  density of such haloes which are capable of producing a significant
  weak-lensing signal with the aperture-mass technique. The contour
  levels are drawn at a halo density of
  $10^{-9.5,-8.5}\,hM_\odot^{-1}$. The different line types indicate
  results for different cosmological models, as indicated. With
  increasing $w\lesssim-0.6$, contours widen, while they shrink as $w$
  increases further.}
\label{fig:9}
\end{figure}

In calculating the number density of haloes of mass $M$ at redshift
$z$ which produce a significant weak-lensing signal,
$N_\mathrm{lens}(M,z)$, we have to take into account that a
signal-to-noise threshold $\mathcal{S}_\mathrm{min}$ for the
weak-lensing signal does not correspond to an equally sharp threshold
in halo mass because of the scatter in the aperture mass which is
caused by the shot noise from the discrete background galaxy positions
and their intrinsic ellipticity distribution. A halo of mass $M$ has a
certain probability $p(M_\mathrm{ap}|M)$ to produce an aperture mass
$M_\mathrm{ap}$, which we model as Gaussian,
\begin{equation}
  p(M_\mathrm{ap}|M)\propto\exp\left\{
    -\frac{[M_\mathrm{ap}-\hat{M}_\mathrm{ap}(M)]^2}
          {2\sigma_\mathrm{M}^2}\right\}\;,
\label{eq:10}
\end{equation}
where $\sigma_\mathrm{M}$ is given by (\ref{eq:9}) and
$\hat{M}_\mathrm{ap}(M)$ is the theoretical expectation for the
aperture mass of a halo with mass $M$. The probability for a halo of
mass $M$ to have an aperture mass above the signal-to-noise threshold
is then
\begin{equation}
  P(\mathcal{S}>\mathcal{S}_\mathrm{min}|M)=
  \frac{1}{2}\,\mathrm{erfc}\left[\frac{\mathcal{S}_\mathrm{min}-
    \hat{\mathcal{S}}(M)}{\sqrt{2}}\right]\;,
\label{eq:11}
\end{equation}
where $\mathrm{erfc}(x)$ is the complementary error function and
$\hat{\mathcal{S}}(M)=\hat{M}_\mathrm{ap}(M)
\sigma_\mathrm{M}^{-1}$. Thus, the number density of significantly
lensing haloes is
\begin{equation}
  N_\mathrm{lens}(M,z)=P(\mathcal{S}>\mathcal{S}_\mathrm{min}|M)\,
  N_\mathrm{halo}(M,z)\;.
\label{eq:12}
\end{equation}
We illustrate $N_\mathrm{lens}(M,z)$ in Fig.~\ref{fig:9}. Here and
below, we set $\mathcal{S}_\mathrm{min}=5$.

The figure shows two sets of contours, inner and outer, whose levels
are $10^{-8.5}$ and $10^{-9.5}\,hM_\odot^{-1}$, respectively. Each set
has four contours for different values of $w$, as indicated in the
figure. All other cosmological parameters are kept fixed,
i.e.~$\Omega_0=0.3$ and $\Omega_\mathrm{Q}=0.7$. The solid contour is
for the $\Lambda$CDM model, for which $w=-1$. The dotted and
short-dashed contours show that the region in the mass-redshift plane
occupied by significantly lensing haloes widens as $w$ increases from
$-1$ to $-0.6$. If $w$ is increased further, this region shrinks
considerably, as the long-dashed contour shows. This illustrates the
competition between the two effects outline above: haloes grow earlier
and are thus more concentrated in dark-energy models with $w>-1$, but
the integrated Sachs-Wolfe effect reduces the power-spectrum
normalisation required by the COBE-DMR data. The maximum extent of the
contours in the mass-redshift plane is reached just before the strong
decrease in $\sigma_8$ near $w\sim-0.6$ illustrated in
Fig.~\ref{fig:3}.

\begin{figure}[ht]
  \includegraphics[width=\hsize]{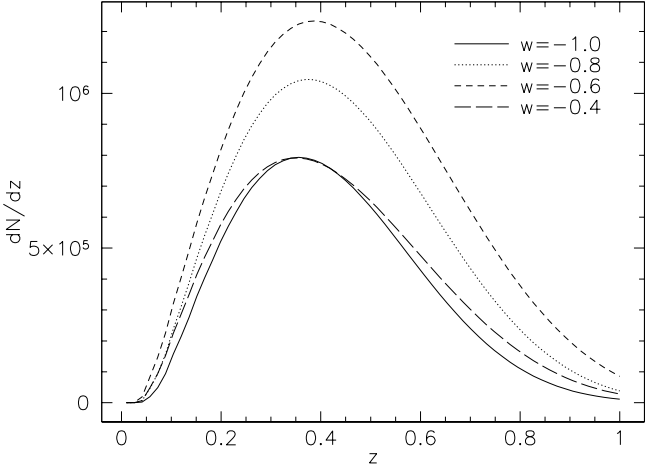}
\caption{The redshift distribution of weak-lensing haloes is shown for
  four different cosmological models, as indicated. As $w$ changes,
  the peak remains near $z\sim0.4$, but the amplitude increases until
  $w\sim-0.6$ and drops for further increasing $w$.}
\label{fig:11}
\end{figure}

Figure~\ref{fig:11} shows the redshift distribution of weak-lensing
haloes for the same four dark energy models used for
Fig.~\ref{fig:9}. The curves in Fig.~\ref{fig:11} are thus integrals
over mass of the distributions in Fig.~\ref{fig:9},
\begin{equation}
  \frac{\d N_\mathrm{lens}(z)}{\d z}=\int_0^\infty\d
  M\,N_\mathrm{lens}(M,z)\;.
\label{eq:13}
\end{equation}

As $w$ increases from $w=-1$ to $w=-0.6$, the amplitude of the
redshift distribution rises by $\sim50\%$, and the redshift
distribution extends towards higher redshift. At $w=-0.4$, the
amplitude drops to the level of the $\Lambda$CDM model ($w=-1$), but
the distribution is somewhat wider. This reflects the fact that the
number of haloes is reduced compared to the models with somewhat lower
$w$ because the power-spectrum normalisation is lower, but the haloes
are more concentrated compared to the $\Lambda$CDM model because of
their earlier formation, allowing them to be significant lenses at
higher redshifts.

\begin{figure}[ht]
  \includegraphics[width=\hsize]{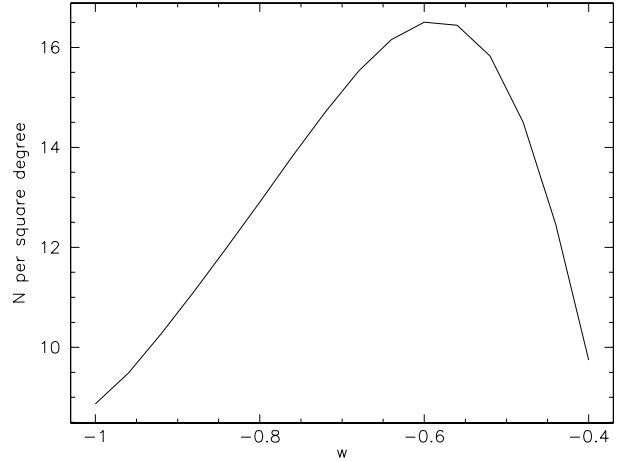}
\caption{Total number of significant weak-lensing haloes per square
  degree as a function of $w$. The curve reaches a peak near
  $w\sim-0.6$, where the halo number is approximately twice that for a
  $\Lambda$CDM model. For $w\gtrsim=-0.6$, the halo number drops
  steeply.}
\label{fig:12}
\end{figure}

Finally, we show in Fig.~\ref{fig:12} the total expected number of
weak-lensing haloes per square degree as a function of $w$. The peak
is reached with $\sim17$ haloes per square degree at $w=-0.6$, which
is almost a factor of two higher than for $\Lambda$CDM models. Note
also that the increase is roughly linear with w up to the maximum.
For larger values of $w$, the halo number drops steeply. However, as
already stressed, these cases are not interesting for cosmology since
they do not produce cosmic acceleration as required by observations.

\section{Summary and Conclusions}

We investigated the expected properties of dark-matter haloes in dark
energy cosmologies. For our purposes, the essential features of such
models are captured describing the dark energy as a density component
with negative pressure $p_\mathrm{Q}=w\rho_\mathrm{Q}$, where $w\ge-1$
is a constant. The dark energy density $\rho_\mathrm{Q}$ is determined
by its present value in units of the critical density,
$\Omega_\mathrm{Q}$. In agreement with results from observations of
the CMB, we focus on models which are spatially flat,
$\Omega_0+\Omega_\mathrm{Q}=1$, have a matter density parameter
$\Omega_0=0.3$, and a Hubble constant of $h=0.7$.

The modified background dynamics in dark-energy models has two
immediate consequences. First, the growth factor is changed, which
determines how structures grow linearly against the expanding
background. In models with fixed parameters $\Omega_0$ and
$\Omega_\mathrm{Q}$, structures form earlier if $w$ is larger. Second,
under equal circumstances, the cosmic volume shrinks as $w$ increases.
Dark energy models thus ``interpolate'' between low-density, spatially
flat models with cosmological constant and low-density, open models.

For our purposes, we can neglect the clustering of the quintessence 
field and assume that the dark-matter power spectrum is given by the 
common CDM spectrum. We 
take the shape parameter to be given by $\Gamma=\Omega_0h$ and
normalise the spectrum such that the COBE-DMR measurements of CMB
fluctuations on large angular scales are reproduced. This implies a
third cosmological consequence. As $w$ increases, the gravitational
potential of matter fluctuations evolves more rapidly along a given
line of sight. Secondary anisotropies in the CMB caused by the
integrated Sachs-Wolfe effect thus grow in amplitude. Keeping the
total fluctuation amplitude fixed to the COBE-DMR data thus requires
the amplitude of the primordial fluctuations to decrease. Expressing
the power-spectrum normalisation by $\sigma_8$, this implies that
$\sigma_8$ must decrease as $w$ increases. The decrease is gentle for
$-1\le w\lesssim-0.6$ and steepens as $w$ increases further.

These findings have two counter-acting effects on the evolution of
dark-matter haloes. First, haloes forming earlier are more
concentrated because their core density reflects the density of the
background universe at their formation time. Since structures form
earlier in dark energy models as $w$ is increased, haloes are expected
to become more concentrated as $w$ grows. Second, the decrease of
$\sigma_8$ with increasing $w$ has the opposite effect on the halo
formation time and indirectly on halo concentration. However,
cosmologically interesting dark-energy equations of state must yield
cosmic acceleration today and require $-1\le w\le -0.6$. Within that
range, the first effect is dominant, and the overall behaviour is
monotonic.

Ideally, extensive, high-resolution numerical simulations would be
necessary for quantifying the net result of these two
effects. However, simple algorithms for calculating halo
concentrations have already been derived from existing numerical
simulations. We used them for our work, assuming that they are also
valid with the modification of Friedmann's equation caused by the
introduction of dark energy instead of a cosmological constant.

We used three different recipes for computing halo
concentrations. Albeit differing in detail, they agree in
concept. Haloes are assigned a formation epoch, essentially requiring
that a certain fraction of the final halo mass has already collapsed
into sufficiently massive progenitors. The characteristic density of
the haloes is then taken to be proportional to the mean background
density of the universe at the halo formation epoch. We showed that
all three recipes lead to the result that haloes are expected to be
increasingly more concentrated as $w$ grows in quintessence models,
showing that the effect of their earlier growth is stronger than the
effect of decreasing $\sigma_8$. This holds for $w\lesssim-0.6$ and
reverses for larger $w$ because the integrated Sachs-Wolfe effect then
requires a steep decrease in $\sigma_8$. The particular recipe for
computing halo concentrations described by Bullock et al.~implies that
haloes should be $\sim50\%$ more concentrated for $w=-0.6$ than for
$w=-1$, where the increase is roughly linear with $w$. 

Finally, we described that halo searches using weak-lensing techniques
are sensitive to halo concentrations. Using the Sheth-Tormen
modification of the Press-Schechter mass function for quantifying the
halo population in mass and redshift, and the aperture mass technique
for quantifying the weak-lensing effects of haloes, we showed that the
expected number density on the sky of haloes causing 5-$\sigma$
weak-lensing detections approximately doubles as $w$ increases from
$-1$ to $-0.6$, where the increase is linear with $w$. Our results
indicate that halo concentrations may be a sensitive probe for the
dark-energy equation of state, and that gravitational lensing may
provide the observational tools for applying that probe. 

Note, however, that we did not allow variations in some cosmological
parameters which may also change the number of weak-lensing haloes. In
particular, an effect may arise from varying the index $n$ of the
dark-matter power spectrum because it directly affects the
determination of $\sigma_8$. On the other hand, our approach here is
to characterise the main effects of dark energy on halo formation and
to propose weak lensing studies as a tool for constraining the dark
energy itself, assuming the main cosmological parameters will be
measured by independent observations like those of the CMB.

Thus, weak lensing turns out to be a powerful tool not only for
mapping the distribution of matter in the Universe, but also for
probing fundamental dark-energy properties. Currently, weak lensing
observations do not allow detailed reconstructions of halo density
profiles (Mellier 2001, Clowe et al.~2000; Mellier \& Van Waerbeke
2001), mainly because of the resolution limit due to the finite number
density of background galaxies.

On the other hand, interesting new perspectives have been opened by
several recent wide-field cosmic-shear studies (Bacon, Refregier \&
Ellis 2000; Bacon, Massey, Refregier, Ellis 2002; Wittman et al.~2000,
Van Waerbeke et al.~2000, Kaiser et al.~2000). Future weak lensing
surveys will cover even larger fields and, thanks to the improved
control of systematic errors, they will allow tighter constraints on
the cosmological parameters. Besides the wide field telescopes which
are currently in their project study phases, we specifically mention
the ``dark matter telescope'' LSST (http://dmtelescop.org, proposed to
scan a 7 square-degree sky field), the VISTA survey
(http://www.vista.ac.uk), and the SNAP satellite
(http://snap.lbl.gov), whose weak lensing survey will cover an area of
300 square degrees, resulting in a very wide field survey with
excellent image quality and depth.

\end{document}